\documentclass[a4paper,12pt]{article}

\usepackage{amssymb,amsmath}
\usepackage[totalwidth=17cm,totalheight=24cm]{geometry}
\usepackage{graphicx}

\newcommand{\R}{{\mathbb R}}

\newcommand{\id}{{\mathbb I}}
\newcommand{\im}{{\rm i\,}}

\newcommand{\e}{{\bf e}}
\newcommand{\m}{{\bf m}}
\newcommand{\A}{{\bf A}}
\newcommand{\B}{{\bf B}}
\newcommand{\F}{{\bf F}}

\newcommand{\be}{\begin{eqnarray}}
\newcommand{\ee}{\end{eqnarray}}

\begin{document}
 \pagestyle{plain}
\title{Dynamics of 3-Forms in Seven Dimensions}
\author{Kirill Krasnov \\ {\it School of Mathematical Sciences, University of Nottingham, NG7 2RD, UK}}

\date{May 2017}
\maketitle

\begin{abstract}\noindent We continue to study a certain dynamical theory of 3-forms in seven dimensions, which can be viewed as a non-linear 7D analog of the 3D Abelian Chern-Simons theory. We perform the $6+1$ split and show that the theory propagates 3 degrees of freedom. We also study the dimensional reduction on $S^3$. We find the resulting theory to be a variant of 4D scalar-tensor theory of gravity. 
\end{abstract}

\section{Introduction}

This work gives a detailed account of the results sketched in \cite{Krasnov:2016wvc}, which further develops the set of ideas put forward in \cite{Herfray:2016azk}, \cite{Herfray:2016std}. This work can be viewed as a continuation of our ongoing effort to unravel implications of the existence of "Deformations of General Relativity" (re)discovered by this author in \cite{Krasnov:2006du}.

From a broader perspective, the motivation for this line of research is the idea that any diffeomorphism invariant theory is either topological or a gravity theory, see below for the implied meaning of the latter. The best known examples of diffeomorphism invariant theories are Schwarz-type topological theories with no propagating degrees of freedom. For instance, these are the theories of differential forms \cite{Schwarz:1978cn}, \cite{Schwarz:1979ae}, as well as their non-Abelian generalisations \cite{Horowitz:1989ng}. Incidentally, gravity in three spacetime dimensions is one of these theories \cite{Witten:1988hc}. 

Another set of examples is given by gravity in 4 or higher spacetime dimensions, and by gravity coupled to matter. For example, in any of the available formulations 4D gravity is a diffeomorphism invariant theory {\it with} propagating degrees of freedom (DOF). Moreover, there are reformulations that depart rather far from the original metric formulation. For example, one can write a rather general set of diffeomorphism invariant gauge theories (for arbitrary gauge group) in 4 dimensions. Any of these theories describes propagating massless gravitons coupled to a set of matter fields \cite{Krasnov:2011hi}. Extrapolating on this set of examples leads to a suspicion that any diffeomorphism invariant theory is either topological (no propagating DOF) or is a gravity theory in the sense of describing propagating gravitons, possibly coupled to matter. 

From this perspective the question of quantum gravity can be reformulated as follows: Are there any diffeomorphism invariant theories with propagating degrees of freedom that make sense quantum mechanically? To answer this question is it first necessary to chart the territory of such theories. The present paper makes a step in this direction and describes a rather unusual such theory - a dynamical theory of 3-forms in 7 dimensions. We will also see that this theory is not so far away from the 4D gravity, as its dimensional reduction to four dimensions will be seen to be an example of a scalar-tensor theory. 

\subsection{Theory of interest}

We consider a theory of 3-forms $\Omega$ on a 7-dimensional manifold $X$. The action is
\be\label{action}
S[\Omega] = \frac{1}{2} \int_X \Omega \wedge d\Omega - \frac{6\lambda}{7} \Omega \wedge {}^*\Omega.
\ee
Here ${}^*\Omega$ is the dual 4-form, computed using the metric on $X$ defined by $\Omega$. Thus, it is a fundamental fact about 3-forms in 7 dimensions that a generic, or in the terminology of \cite{Hitchin:2001rw} {\it stable} 3-form defines a metric via the formula
\be\label{metric}
g_\Omega(\xi,\eta) {\rm Vol}_\Omega = -\frac{1}{6} i_\xi \Omega \, i_\eta \Omega \, \Omega.
\ee
Here $\xi,\eta$ are vector fields, and $i_\xi$ is the operation of the insertion of a vector field into a form (i.e. interior product). The forms on the right-hand-side of (\ref{metric}) are multiplied using the wedge product, whose symbol is omitted for brevity. With the basic field of the theory defining the metric, the theory (\ref{action}) can be viewed as a gravity theory in the sense that its field equations will constrain the metrics that arise. 

It is a classical fact that for real $\Omega$ the metric that arises via (\ref{metric}) is either Riemannian or of signature $(4,3)$. We will consider the real 3-forms $\Omega$ that give rise to Riemannian signature metrics. Thus, our theory will be analogous to Euclidean signature gravity. In Euclidean signature gravity one can meaningfully discuss quantum gravity and its problems by attempting to compute the Euclidean path integral. Similar questions are possible for the theory (\ref{action}), but we will only make some preparatory steps in the direction of these questions in the present paper. For some additional comments on the problem of quantisation of (\ref{action}) see the last section. 

The last term in (\ref{action}) is of homogeneity degree $7/3$ in $\Omega$, and so the Euler-Lagrange equations arising by minimising the above action are
\be\label{feqs}
d\Omega = \lambda {}^*\Omega.
\ee
Thus, the 3-forms $\Omega$ that are the critical points of the above functional give $X$ the structure of a manifold with weak holonomy $G_2$, also called nearly parallel $G_2$ structure in \cite{Friedrich:1995dp}. Such manifolds are singled out by the fact that they admit Killing spinors. In fact, as the reference \cite{Friedrich:1995dp} shows, see also Appendix A of \cite{Becker:2014rea} for a physicist-friendly discussion, a stable 3-form $\Omega$ can be parametrised by a pair (metric, unit spinor). The equation (\ref{feqs}) then receives the interpretation of the equation stating that the spinor is a Killing spinor. This in particular implies that the metric is Einstein of constant scalar curvature $21/8 \lambda^2$, see proposition 3.10 from \cite{Friedrich:1995dp}. Note that the scalar curvature is necessarily positive. 

We also note that (\ref{feqs}) imply in particular that the dual form ${}^*\Omega$ is closed
\be\label{d-feqs}
d{}^*\Omega=0.
\ee

\bigskip
\noindent{\bf Remarks:}
Equations (\ref{feqs}) were also derived in \cite{Hitchin:2001rw} from a constrained variational principle. The basic object was taken to be an exact 4-form $\rho=d\gamma$, and the quantity minimised was the associated volume. The variation was subject to constraint that $\gamma\wedge d\gamma$ is fixed. It is clear that our variational principle (\ref{action}) is very much related. Indeed, the constraint in the case of \cite{Hitchin:2001rw} can be imposed with a Lagrange multiplier term added to the action. Then, with the identification $\gamma=\Omega$ we get both terms in (\ref{action}). Our variational principle gives (\ref{feqs}) in a more direct fashion, without use of any Lagrange multipliers. 

The theory defined by just the first term of our action (\ref{action}) was already studied in the literature, see \cite{Gerasimov:2004yx}. In particular, this reference discussed the fact that this theory is topological. The action (\ref{action}) also appears in reference \cite{Nekrasov:2004vv}, formula (29), in a generalised version that includes 1- and 5-forms.  

\subsection{Examples of critical points}

There are many known examples of solutions of (\ref{feqs}), see e.g. \cite{Friedrich:1995dp}. Be give a brief account. 

The examples can be classified by the number of Killing spinors that exist. There are two known examples with the number of Killing spinors equal to three. The principal example is the 7-dimensional sphere $S^7$ with the standard metric, i.e. $S^7={\rm SO}(8)/{\rm SO}(7)$. The other example is the space ${\rm SU}(3)/{\rm U}(1)$. 

Examples with 2 Killing spinors can be obtained as principal $S^1$ bundles over a 6-dimensional K\"ahler-Einstein manifolds with positive scalar curvature. A list of possible 6-dimensional manifolds appearing in this context is given in a table on page 13 of \cite{Friedrich:1995dp}.

Examples with 1 Killing spinor are as follows. First, there is the squashed 7-sphere, which is the second Einstein metric that can be put on the principal ${\rm SU}(2)$ bundle over $S^4$, with the first of these metrics being the round $S^7$. Second, there are spaces ${\rm SU}(3)/ S^1_{k,l}$, where the embedding of $S^1$ into ${\rm SU}(3)$ is given by 
\be
{\rm U}(1)\ni z \to {\rm diag}(z^k, z^l, z^{-(k+1)}) \in {\rm SU}(3).
\ee
Each of these spaces admits two homogeneous Einstein metrics. When $(k,l)=(1,1)$ one of these metrics has 3 Killing spinors, and the other has one. For $(k,l)\not=(1,1)$ both of these metrics have just one Killing spinor. The last example is the space ${\rm SO}(5)/{\rm SO}(3)$, where ${\rm SO}(3)$ is embedded into ${\rm SO}(5)$ via the so-called principal embedding. 

It is also worth mentioning that the cone over a metric defined by one of the solutions to (\ref{feqs}) is an 8-dimensional manifold of holonomy ${\rm Spin}(7)$. 

\subsection{Results}

Results obtained in the main text are as follows. First, in Section \ref{sec:6+1} we perform the Hamiltonian $6+1$ decomposition of (\ref{action}). The unreduced phase space is easily seen to be the space of 3-forms in 6 dimensions, and so is of dimension 20. We will see that there are 7 first class constraints, and the constraint algebra is just the algebra of 7D diffeomorphisms. This immediately implies the count of propagating degrees of freedom $(20-2\times 7)/2 = 3$. While this result seems to follow with almost no analysis, it takes some work to explicitly exhibit the constraints.

As we have already mentioned above, the critical points of (\ref{action}) have the interpretation of Einstein 7-manifolds with positive scalar curvature and with a Killing spinor. The condition of admitting a Killing spinor implies the Einstein property, but is much stronger. Our count of the number of propagating degrees of freedom confirms this. Einstein's theory in 7 dimensions has $28-2\times 7=14$ propagating DOF. The theory (\ref{action}) is much more constraining in particular in the sense that the number of propagating DOF is much smaller. 

The other result of this paper is characterisation of the theory (\ref{action}) in terms of its dimensional reduction to 4D, by reducing on $S^3$. Thus, in Section \ref{sec:4+3} we assume that ${\rm SU}(2)$ acts on our 7D manifold without fixed points, and that the 3-form $\Omega$ is invariant under this action. We parametrise the general such 3-form in terms of some 4D data, and show that the dimensionally reduced theory is a particular scalar-tensor theory. This gives an interpretation to the 3 degrees of freedom found in our Hamiltonian analysis after the dimensional reduction - 2 of these degrees of freedom are those of a 4D graviton, and the other one is that of a scalar. We also describe explicitly in Section \ref{sec:spheres} the two solutions of (\ref{feqs}) that are $S^3$ fibrations over $S^4$. 

Another new and potentially interesting in its own right result of our work is the fact that the dimensional reduction of the topological $\Omega d\Omega$ theory to 4D is the well-known ${\rm SU}(2)$ BF theory with the cosmological term. The letter theory is of course also topological. To the best of our knowledge, this is the first example of a Kaluza-Klein-type relation between a Schwarz-type topological theory of differential forms in higher dimensions, and a non-Abelian topological theory of BF-type in lower dimensions. Given that the quantisation of the letter theories is reasonably well-understood, in particular in terms of state sum models, it is an interesting question if state sum model quantisation can also be carried out for the theories of the former type. 

\subsection{7-Dimensional origin of the Urbantke formula}

Another aspect of 7-dimensional geometry that is worth emphasising in the Introduction is the fact that the formula (\ref{metric}) for the metric as determined by a 3-form provides an explanation to the so-called Urbantke formula \cite{Urb}. 

It is well-known that in 4 dimensions a triple of 2-forms satisfying the condition that the matrix of their wedge products has a definite sign determines a Riemannian signature metric. Urbantke gave an explicit expression for this metric. Thus, let $B^i, i=1,2,3$ be a triple of 2-forms. The Urbantke metric is then determined via
\be\label{urb-metric}
g_B(\xi,\eta) {\rm Vol}_B = \frac{1}{6} \epsilon^{ijk} i_\xi B^i i_\eta B^j B^k.
\ee 
Here ${\rm Vol}_B$ is the volume form for the metric $g_B$. 

Let us now consider a 7D manifold $X$ that has the structure of an $S^3$ bundle over a 4-dimensional base $M$. As it is made clear by our explicit computation leading to (\ref{met-1}), the metric induced on $M$ by (\ref{metric}) as determined by an ${\rm SU}(2)$-invariant 3-form
\be\label{omega-intr}
\Omega = -2{\rm Tr}\left( \frac{1}{3} W^3 + WB\right)
\ee
is just the Urbantke metric (\ref{urb-metric}). Here $W=g^{-1} dg + g^{-1}{\bf A} g, B=g^{-1} {\bf B} g$ are Lie algebra valued 1- and 2-forms in the total space of an ${\rm SU}(2)$ bundle over $M$, and ${\bf A}, {\bf B}$ are Lie algebra valued 1- and 2-forms on $M$. 

So, the triple of 2-forms $B^i$ that determines a 4D metric on $M$ according to (\ref{urb-metric}) can be viewed as vertical-horisontal-horisonal components of a 3-form (\ref{omega-intr}) in the 7-dimensional space obtained by attaching 3D fibers to every point of $M$. The Urbantke formula (\ref{urb-metric}) is then just a special case of the 7D formula (\ref{metric}). 

We find the described embedding of the 4D space $M$ with a triple of 2-forms on $M$ into a 7D space with a 3-form as a conceptually clear explanation for why 2-forms in 4D determine a metric. 

The encoding of a 4D metric into a triple of 2-forms plays the key role in all formulations of 4D General Relativity related to Plebanski formalism \cite{Plebanski:1977zz}, see also \cite{Krasnov:2009pu}. In particular, this encoding is the essential feature of the construction of "Deformations of GR" \cite{Krasnov:2006du}. The described above 7D explanation of the Urbantke formula makes one suspect that also 4D GR in Plebanski formulation should have some 7D origin. This paper can be viewed as a step towards establishing this 7D perspective on GR.

\section{Hamiltonian analysis}
\label{sec:6+1}

In this section we carry out the Hamiltonian reduction of the theory (\ref{action}) and confirm the count of the number of physical degrees of freedom as stated in the Introduction.

\subsection{Topological part}

Let us carry out the $6+1$ decomposition of the theory (\ref{action}), assuming $X=\R\times \Sigma$. We write
\be\label{omega-6+1}
\Omega = dt\wedge B + C,
\ee
where $B$ and $C$ are a 2- and 3-forms on the 6-dimensional slice. We then have
\be
d\Omega = dt\wedge (\dot{C} - dB) + dC,
\ee
and
\be
\int \Omega d\Omega = \int dt\int_\Sigma \left( -C\dot{C} + 2B dC\right),
\ee
where we neglected total derivative terms in the integration over $\Sigma$. This makes it clear that the theory given just by the first term in (\ref{action}) is topological. Its phase space is the space of 3-forms $C$ on $\Sigma$, which is 20 dimensional. The Hamiltonian is a constraint, with $B$ field playing the role of the Lagrange multiplier imposing $dC=0$. The transformation that this constraint generates on the phase space is 
\be\label{delta-C}
\delta C=dB.
\ee
Constraints together with gauge transformations they generate reduce the phase space to a finite-dimensional one - the reduced phase space is the third cohomology $H^3(\Sigma)$ of the "spatial" slice $\Sigma$. 

Of course, the presence of the second term in (\ref{action}) renders the theory not topological. Our goal is to understand the Hamiltonian picture in this case. 

\subsection{Computation of the metric}

To compute the volume form part of the Lagrangian, we compute the metric defined by $\Omega$, in the parametrisation of $\Omega$ given by (\ref{omega-6+1}). The metric is defined via (\ref{metric}). 
A straightforward computation gives
\be\label{metric6-2}
i_\xi \Omega i_\eta \Omega \, \Omega / dt = i_\xi dt  \, i_\eta dt B^3 
+ \frac{3}{2} \left( i_\xi dt \, BB i_\eta C + i_\eta dt \, BB i_\xi C\right) + 3B i_\xi C i_\eta C.
\ee
The first and last terms on the right-hand-side are the time-time and space-space components of the metric respectively. The middle term is the off-diagonal part of the metric. It can be removed by rewriting the metric as
\be\label{metric6-1}
i_\xi \Omega i_\eta \Omega \, \Omega / dt = i_\xi( dt + \alpha)\, i_\eta(dt+\alpha) B^3 + 3B i_\xi C i_\eta C - i_\xi\alpha \, i_\eta\alpha \, B^3
\ee
where the 1-form $\alpha$ is determined from
\be\label{alpha}
i_\xi \alpha = \frac{3}{2} \frac{BB i_\xi C}{BBB}.
\ee
It is assumed here that the 2-form $B$ is non-degenerate, so that $BBB$ is non-zero. The presence of $\alpha$ in the first term on the right-hand-side of (\ref{metric6-1}) has clear geometrical meaning. Indeed, the 7-dimensional metric determines a connection in the $\R$ bundle over $\Sigma$, by declaring the vectors orthogonal to vertical to be horizontal. The corresponding connection 1-form is $\alpha$. 

Let us also interpret the arising metric on $\Sigma$. We first note that the 1-form $\alpha$ given by (\ref{alpha}) also arises when we rewrite
\be\label{omega6-alt}
\Omega = (dt+\alpha) B + \tilde{C},\qquad \tilde{C} = C - \alpha B,
\ee
and impose the equation
\be
\tilde{C} B=0.
\ee
Indeed, the equation we are solving reads
\be
CB = \alpha BB.
\ee
Let us insert an arbitrary vector field into 5-forms on both sides
\be
i_\xi C B - C i_\xi B = i_\xi \alpha BB - 2\alpha i_\xi B B,
\ee
and multiply with another $B$. We get, after some simple identities
\be\label{CBB}
\frac{1}{2} i_\xi C BB = \frac{1}{3} i_\xi \alpha BBB,
\ee
whose solution is clearly (\ref{alpha}). We work under assumption that the 2-form $B$ on $\Sigma$ is non-degenerate (has non-zero determinant), or equivalently, that the top form $BBB$ is nowhere vanishing. 

But now that $\Omega$ is written in the alternative form (\ref{omega6-alt}) with $\tilde{C}B=0$ we see that there is no off-diagonal term in the metric (\ref{metric6-2}). The metric takes the block-diagonal form (\ref{metric6-1})
\be\label{metric6-3}
i_\xi \Omega i_\eta \Omega \, \Omega / dt = i_\xi( dt + \alpha)\, i_\eta(dt+\alpha) B^3 + 3B i_\xi \tilde{C} i_\eta \tilde{C}. 
\ee

\subsection{Interpretation of the metric on the base}

We now want to show that the metric on $\Sigma$ given by the second term on the right-hand-side of (\ref{metric6-3}) has a simple interpretation. Recall \cite{Hitchin:2000jd} that a 3-form on 6 dimensions defines an endomorphism of the tangent bundle that squares to plus or minus identity. The sign depends on the ${\rm GL}(6,\R)$ orbit to which the 3-form belongs - there are exactly two orbits distinguished by this sign. 

The endomorphism is constructed explicitly as follows. It is convenient to choose some volume form $v$ on $\Sigma$, the end result will only depend on the orientation of $v$. We first define an endomorphism $K_{\tilde{C}}$ that squares to a multiple of the identity, and then rescale. Let us define the action of $K_{\tilde{C}}$ on a 1-form $\eta$ as follows
\be
i_\xi K_{\tilde{C}}(\eta) := \eta\, i_\xi \tilde{C} \, \tilde{C}/ v.
\ee
We emphasise that an arbitrary top form $v$ can be used in the denominator on the right-hand-side. On the left $K_{\tilde{C}}(\eta)$ is the 1-form that is the result of the action of $K_{\tilde{C}}$ on $\eta$. It can be verified that $K_{\tilde{C}}^2 = \lambda_{\tilde{C}} \id$ and so ${\rm Tr}(K_{\tilde{C}}^2) = 6\lambda_{\tilde{C}}$. It is convenient to introduce the notation
\be
{\rm Vol}_{\tilde{C}} := \sqrt{\frac{\pm{\rm Tr}(K^2)}{6}} v.
\ee
Note that this is a well-defined volume form that depends only on the orientation of the auxiliary volume form $v$ used in the definition. We then define
\be
J_{\tilde{C}} := \frac{v}{{\rm Vol}_{\tilde{C}}} K_{\tilde{C}}.
\ee
The endomorphism $J_{\tilde{C}}$ depend on the volume form used in the construction of $K_{\tilde{C}}$ only via the orientation $v$ defines. It squares to plus or minus identity, depending on the sign of $\lambda_{\tilde{C}}$, equivalently on the type of the 3-form $\tilde{C}$. The endomorphism $J_{\tilde{C}}$ is then either an almost complex structure when $J_{\tilde{C}}^2=-\id$, or an almost para-complex structure when $J_{\tilde{C}}^2=\id$. For definiteness, below we shall assume that we have the case of almost complex structure.

If, in addition to $\tilde{C}$, we also have a 2-form $B$ in our disposal, we can define a metric. This is not surprising since $\tilde{C},B$ together form, see (\ref{omega6-alt}), a 3-form in one dimension higher, and this defines a metric. But the 6D metric that this 7D metric would induce can also be understood purely in 6D terms. Indeed, we can form a 6D tensor by inserting $J_{\tilde{C}}$ in one of the slots of $B$. It is then easy to check that the condition $\tilde{C}B=0$ guarantees that this tensor is symmetric. Indeed, we have
\be
2B(\xi,J_{\tilde{C}}\eta) = i_\eta J_{\tilde{C}}(i_\xi B) = i_\xi B i_\eta \tilde{C}\, \tilde{C} / {\rm Vol}_{\tilde{C}}.
\ee
We then use
\be
0= i_\xi (B i_\eta \tilde{C}\, \tilde{C}) = i_\xi B i_\eta \tilde{C} \, \tilde{C} + B i_\xi i_\eta \tilde{C} \, \tilde{C} + B i_\eta \tilde{C} \, i_\xi \tilde{C}
\ee
to see that when $\tilde{C}B=0$ we have
\be
2B(\xi,J_{\tilde{C}}\eta) = - B i_\xi \tilde{C} \, i_\eta \tilde{C} / {\rm Vol}_{\tilde{C}},
\ee
which is explicitly $\xi,\eta$ symmetric. We now define
\be\label{metric-B}
g_B(\xi,\eta) := B(\xi,J_{\tilde{C}} \eta).
\ee
The metric (\ref{metric6-3}) is then rewritten as 
\be\label{metric6-4}
\frac{1}{6} i_\xi \Omega i_\eta \Omega \, \Omega / dt = i_\xi( dt + \alpha)\, i_\eta(dt+\alpha) {\rm Vol}_B -   g_B(\xi,\eta){\rm Vol}_{\tilde{C}},
\ee
where we introduced
\be
{\rm Vol}_B := \frac{1}{6} B^3.
\ee
Note that for the case of $\tilde{C}$ of negative type, which corresponds to $\tilde{C}$ defining an almost complex structure, the natural orientations of ${\rm Vol}_B$ and ${\rm Vol}_{\tilde{C}}$ are opposite of each other, which explains the relative minus sign in the above formula. This fact is easiest verified by taking $\tilde{C} = \alpha_1\alpha_2\alpha_3 + \bar{\alpha_1} \bar{\alpha_2} \bar{\alpha_3}$, and $v= (\im)^3  \alpha_1\alpha_2\alpha_3\bar{\alpha_1} \bar{\alpha_2} \bar{\alpha_3}$. In this case $K_{\tilde{C}}(\alpha_1) =\im \alpha_1$, so the forms $\alpha_{1,2,3}$ are holomorhic. We also have ${\rm Vol}_{\tilde{C}}=v$. We can then take $B=\im(\alpha_1\bar{\alpha}_1+\alpha_2\bar{\alpha}_2+\alpha_3\bar{\alpha}_3)$ as the 2-form that gives an all plus signature metric via (\ref{metric-B}). It is then clear that ${\rm Vol}_B = (\im)^3 \alpha_1\bar{\alpha}_1\alpha_2\bar{\alpha}_2\alpha_3\bar{\alpha}_3= -{\rm Vol}_{\tilde{C}}$. Thus, in the case of $\tilde{C}$ of negative type, we have a Riemannian metric on the right-hand-side of (\ref{metric6-4}). In the opposite case of positive type $\tilde{C}$ we will get a metric of signature $(3,3)$ for $g_B$, and thus the metric of signature $(4,3)$ for $g_\Omega$. 

\subsection{Computation of the volume form}

We can now compute the determinant of the metric (\ref{metric6-4}). The volume form for $\Omega$ is obtained as the $1/9$ power of this determinant. For the determinant of $g_B$ we have 
\be
{\rm det}(g_B) = {\rm det}(B) {\rm det}(J_{\tilde{C}}) = {\rm det}(B) = {\rm Pf}(B)^2.
\ee
A simple computation then gives
\be\label{vol-omega}
{\rm Vol}_\Omega = dt \left( {\rm Vol}_B\right)^{1/3} \left( {\rm Vol}_{\tilde{C}}\right)^{2/3}.
\ee 

\subsection{Action in the Hamiltonian form}

We can now write the full action in the Hamiltonian form. The potential term in the action can be written as a multiple of the volume form 
\be
S[\Omega] = \frac{1}{2} \int_X \Omega d\Omega + 6\lambda {\rm Vol}_\Omega.
\ee
Collecting the results above we have
\be\label{act-ham}
S[\Omega]=\int dt\int_\Sigma -\frac{1}{2} C\dot{C} + BdC + 3\lambda \left( {\rm Vol}_B\right)^{1/3} \left( {\rm Vol}_{\tilde{C}}\right)^{2/3}.
\ee

\subsection{Constraint equation}

Varying (\ref{act-ham}) with respect to the 2-form field $B$ with obtain an equation without time derivatives -- a constraint
\be\label{constr-1}
dC \delta B + \lambda \left( {\rm Vol}_B\right)^{1/3} \left( {\rm Vol}_{\tilde{C}}\right)^{2/3}
\left( \frac{BB \delta B}{2 {\rm Vol}_B} - 2\frac{\delta (\alpha B) \hat{\tilde{C}}}{{\rm Vol}_{\tilde{C}}}\right)=0.
\ee 
Here we have used
\be
2{\rm Vol}_{\tilde{C}} = \tilde{C} \hat{\tilde{C}},
\ee
which thus implies
\be
\delta {\rm Vol}_{\tilde{C}} = \delta \tilde{C} \hat{\tilde{C}}
\ee
because $\hat{\tilde{C}}$ is function of degree of homogeneity 2 in $\tilde{C}$. We also used the fact that under variation of $B$ we have $\delta\tilde{C}=\delta(\alpha B)$. 

The second term in (\ref{constr-1}) can be simplified by noticing that 
\be\label{B-hat-C}
B \hat{\tilde{C}} =0,
\ee
which follows from $B\tilde{C}=0$. Indeed, let us act on all indices of the 5-form $B\tilde{C}$ with the endomorphism $J_{\tilde{C}}$. The action on $B$ gives $B(J_{\tilde{C}} \cdot, J_{\tilde{C}} \cdot)=-B(\cdot,\cdot)$. The action on $\tilde{C}$ gives by definition the form $\hat{\tilde{C}}$. So, we obtain (\ref{B-hat-C}). Thus, $\delta(\alpha B) \hat{\tilde{C}}= \alpha \hat{\tilde{C}} \delta B$, and we can write the constraint as
\be\label{constr-2}
dC  + \lambda \left( {\rm Vol}_B\right)^{1/3} \left( {\rm Vol}_{\tilde{C}}\right)^{2/3}
\left( \frac{BB }{2 {\rm Vol}_B} - 2\frac{\alpha  \hat{\tilde{C}}}{{\rm Vol}_{\tilde{C}}}\right)=0.
\ee 

\subsection{The evolution equation}

Let us now vary the action with respect to $C$. We get
\be\label{evol}
\dot{C} - dB = 2\lambda \frac{\left( {\rm Vol}_B\right)^{1/3}}{\left( {\rm Vol}_{\tilde{C}}\right)^{1/3}} \hat{\tilde{C}}.
\ee
Here, on the right-hand-side, we have replaced $\delta{\tilde{C}}=\delta(C - \alpha B) = \delta C - \delta\alpha B$ with $\delta C$ because the $\delta\alpha B$ term, when multiplied with $\hat{\tilde{C}}$ gives zero because of (\ref{B-hat-C}). 

\subsection{The expression for ${}^*\Omega$}

We now note that the two equations (\ref{constr-2}) and (\ref{evol}) we obtained together imply an expression for ${}^*\Omega$. Indeed, we know that both equations together are equivalent to $d\Omega = \lambda {}^*\Omega$. We can rewrite the constraint and evolution equations as
\be
dC  = \lambda \left( {\rm Vol}_B\right)^{1/3} \left( {\rm Vol}_{\tilde{C}}\right)^{2/3} \left( 2\frac{\alpha  \hat{\tilde{C}}}{{\rm Vol}_{\tilde{C}}}- \frac{BB }{2 {\rm Vol}_B} \right), \\ \nonumber
dt (\dot{C} - dB) = \lambda \left( {\rm Vol}_B\right)^{1/3} \left( {\rm Vol}_{\tilde{C}}\right)^{2/3} \left( 2\frac{dt \hat{\tilde{C}}}{{\rm Vol}_{\tilde{C}}}\right).
\ee
Adding them together and comparing with $d\Omega = \lambda {}^*\Omega$ we read off
\be
{}^* \Omega = \left( {\rm Vol}_B\right)^{1/3} \left( {\rm Vol}_{\tilde{C}}\right)^{2/3} \left( 2\frac{(dt+\alpha)  \hat{\tilde{C}}}{{\rm Vol}_{\tilde{C}}}- \frac{BB }{2 {\rm Vol}_B} \right).
\ee
As a check, we note that we get the correct relation $-7 {\rm Vol}_\Omega = \Omega {}^*\Omega$, with 
the 7D volume form given by (\ref{vol-omega}). 

\subsection{Some consequences of the constraint equation}

We now derive some consequences of (\ref{constr-2}). First, let us multiply this equation with $i_\xi C$, for an arbitrary vector field $\xi$. To see what the result is, let us simplify the second term in the brackets. We have
\be
i_\xi C \alpha \hat{\tilde{C}} = i_\xi (\tilde{C} +\alpha B) \alpha \hat{\tilde{C}} =  i_\xi \tilde{C} \alpha \hat{\tilde{C}}.
\ee
The last equality arises because $i_\xi (\alpha B)=i_\xi \alpha B - \alpha i_\xi B$, and this vanishes when multiplied with $\alpha \hat{\tilde{C}}$ because of (\ref{B-hat-C}). We then have $i_\xi \tilde{C} \hat{\tilde{C}} = i_\xi {\rm Vol}_{\tilde{C}}$, and thus $i_\xi \tilde{C} \alpha \hat{\tilde{C}} = \alpha i_\xi {\rm Vol}_{\tilde{C}} = i_\xi \alpha {\rm Vol}_{\tilde{C}}$. Therefore, in view of (\ref{CBB}), the term in the brackets in (\ref{constr-2}), multiplied with $i_\xi C$ becomes $2i_\xi \alpha - 2i_\xi \alpha=0$. Thus, we get
\be\label{diff-constr}
i_\xi C dC=0,
\ee
as a consequence of (\ref{constr-2}). This is independent of $B$, and is a true constraint on the phase space, which is the space of 3-forms on $\Sigma$. 

\subsection{Interpretation of the constraints}

The interpretation of the constraint equation (\ref{constr-2}) is that it should be used to determine $B$, which is then to be inserted into the evolution equations (\ref{evol}). However, the fact that (\ref{constr-2}) implies (\ref{diff-constr}) means that we cannot solve for all the components of $B$. Some arbitrary functions will remain in $B$, and the presence of these arbitrary functions is reflected in particular in the existence of the constraints (\ref{diff-constr}). 

From the form of (\ref{constr-2}) it is also clear that the scale of $B$ cannot be solved for. Indeed, the constraint (\ref{constr-2}) is homogeneity degree zero in $B$. This is checked as follows. First, the 1-form $\alpha$ is of degree $-1$ in $B$, as is seen from (\ref{CBB}). Second, because $\tilde{C}=C-\alpha B$ it is clear that $\tilde{C}$ is of homogeneity degree zero. This count makes it clear that all the terms in (\ref{constr-2}) are of homogeneity degree zero. Thus, the scale of $B$ can never be fixed from this constraint equation. This signals presence of one more free function in $B$, in addition to those present due to (\ref{diff-constr}). 

Thus, it is to be expected that 7 of the 15 components of $B$ cannot be solved for from (\ref{constr-2}). This is not surprising, because the theory is diffeomorphism invariant, so we do expect at least 7 arbitrary functions entering the evolution equations. 

So, it is natural to expect that (\ref{diff-constr}) are just the 6 spatial diffeomorphism constraints. And indeed, the induced transformation of $C$ is 
\be
\delta_\xi C = \{ C, \int i_\xi C dC \} = i_\xi dC + d i_\xi C = {\cal L}_\xi C,
\ee
which confirms the interpretation of (\ref{diff-constr}) as diffeomorphism constraints. 

To deduce the interpretation of the last constraint, related to the inability to solve for the scale of $B$, let us multiply (\ref{constr-2}) with $B$. We get
\be\label{h-constr}
BdC + 3\lambda {\rm Vol}_\Omega/dt = 0,
\ee
where we have used (\ref{vol-omega}). We can multiply this constraint with an arbitrary smearing function $\mu$ and take the Poisson bracket with $C$ to see what it will generate. The result is
\be
\delta_\mu C = d(\mu B) - 3\lambda\mu \frac{\partial }{\partial C} \left( \frac{{\rm Vol}_\Omega}{dt}\right).
\ee
On the other hand, in view of (\ref{evol}), this is just
\be
\delta_\mu C = d(\mu B) + \mu (\dot{C}-dB) \equiv {\cal L}_{\mu \partial/\partial t} \Omega,
\ee
where it is understood that in the Lie derivative on the right-hand-side of the last equality the projection onto the slice $\Sigma$ is taken. This establishes the interpretation of (\ref{h-constr}) as the Hamiltonian constraint, i.e. as the constraint generating the timelike diffeomorphisms. 

Thus, the constraints (\ref{diff-constr}), (\ref{h-constr}) generate diffeomorphisms. Therefore, they have the usual algebra. In particular, the algebra closes. 

This allows us to do the count of the number of physical degrees of freedom. The dimension of the unreduced phase space is 20. We have 7 constraints. The constraints act on the constraint surface, with the space transverse to the orbits being of codimension 7. Thus, the dimension of the reduced phase space is $20-7-7=6$, which means 3 physical DOF. Of course, this result could have been anticipated from the start, prior to any analysis. However, we believe it's important to exhibit the constraints explicitly, which is what we have done in this section. 

\section{Dimensional reduction to 4D}
\label{sec:4+3}

The aim of this section is to perform the dimensional reduction of the theory (\ref{action}) to 4D, assuming that the 3-form $\Omega$ is invariant under the action of ${\rm SU}(2)$ on $X$. Thus, we assume that ${\rm SU}(2)$ acts on our 7-dimensional manifold $X$ without fixed points, so that the manifold takes the form of a principal ${\rm SU}(2)$ bundle over a 4-dimensional base $M$. The form $\Omega$ can then be parametrised in terms of some data on the base $M$. 

\subsection{Parametrisation}

We choose to parametrise the 3-form in the following way
\be\label{C}
\Omega = -2 {\rm Tr}\left( \frac{1}{3} \phi^3\, W^3 + \phi W B\right) + c.
\ee
Here $\phi$ is a scalar field, $g\in{\rm SU}(2)$, $c$ is a 3-form on the base and 
\be
W= g^{-1} dg+A, \qquad A = g^{-1} \A g, \qquad B=g^{-1} \B g
\ee
are a connection on the total space of the bundle and the lifts of the Lie algebra valued 1-form $\A$ and 2-form $\B$ on the base to the total space of the bundle. The objects $W,B$ are Lie algebra valued forms in the total space of the bundle, which here means that they are forms with values in the space of $2\times 2$ anti-Hermitian matrices. 

The total number of fields described by $\phi,{\bf A}, {\bf B}, c$ is $1+12+18+4=35$, which is the total number of components of a 3-form in 7 dimensions. The form (\ref{C}) is invariant under (global) right ${\rm SU}(2)$ transformations. Diffeomorphisms along the fiber are realised as gauge transformations. 

A simple computation gives
\be
d\Omega= - 2 \,{\rm Tr} \Big( \phi^2 d\phi W^3 +( \phi^3 F+ \phi B) W^2 \\ \nonumber 
+( d\phi B + \phi d_A B)W + \phi FB \big) + dc.
\ee
Here $F=g^{-1} \F g$ is the lift to the total space of the curvature $\F=d\A+\A^2$, and $d_A B = g^{-1} (d{\bf B}+{\bf A}{\bf B}-{\bf B}{\bf A}) g$ is the lift to the bundle of the covariant derivative of Lie algebra-valued 2-form ${\bf B}$ with respect to the connection $\bf A$. 

\subsection{The dimensional reduction of $\Omega d\Omega$ and 4D BF theory}

Another simple computation gives
\be\label{cdc}
\frac{1}{2} \int_{X} \Omega d\Omega = \int_{{\rm SU}(2)} -\frac{2}{3} {\rm Tr}(\m^3) \\ \nonumber \times
\int_M -2\, {\rm Tr}(\phi^4 {\bf B}{\bf F} + (\phi^2/2) {\bf B}{\bf B}) + \phi^3 dc.
\ee
In deriving this result we have used the following trace identities
\be\label{trace-identities}
(-2{\rm Tr}(W^2 a))(-2{\rm Tr}(W b)) = -\frac{2}{3} {\rm Tr}(W^3) ( -2{\rm Tr}(ab)), \\ \nonumber
(-2{\rm Tr}(W a))(-2{\rm Tr}(W b)) (-2{\rm Tr}(W c))=-\frac{2}{3} {\rm Tr}(W^3) ( -4{\rm Tr}(abc)).
\ee

Thus, the result of dimensional reduction of $\Omega d\Omega$ is the Lagrangian of a topological field theory in 4 dimensions. It is the so-called BF theory with the cosmological constant, appended with an extra term describing the 3-form field $c$. Its Euler-Lagrange equations state 
\be
d\phi =0, \quad dc =0, \quad \phi^2\F=-\B.
\ee

\subsection{Topological symmetry}

The theory (\ref{cdc}) is topological because it has a very large gauge symmetry. Let us see how this symmetry arises from the symmetry of the original 7-dimensional Lagrangian. It is clear that the first term in (\ref{action}) is invariant under $\delta \Omega=dH$, where $H$ is a 2-form. Assuming that $H$ is an invariant 3-form (with respect to the action in the fibers) we can write it as
\be
H = {\rm Tr}\left( g^{-1} \psi g \m^2 + g^{-1} \eta g \m \right) + h.
\ee
Here $\psi$ is a Lie-algebra valued scalar, and $\eta$ is a Lie algebra valued 1-form on the base. Its exterior derivative is given by
\be
dH = {\rm Tr}\left( g^{-1} (d \psi -\eta) g\, \m^2 + g^{-1} d\eta g\, \m\right) + dh.
\ee
Comparing with (\ref{C}) expanded in powers of $\m$ we immediately read off the transformation laws for this gauge symmetry
\be
\delta \phi =0, \quad \delta c+\delta\left( \frac{1}{3} \phi^3 A^3 + \phi AB\right)=dh,\quad \phi^3 \delta \A = d\psi -\eta, \quad \phi \delta(\B+\phi^2 \A^2) = d\eta.
\ee
We can rewrite the last two laws in a more familiar form as follows. First, we change the exterior derivative by the covariant derivative and redefine $\eta$
\be\label{delta-A}
\phi^3 \delta \A = d_\A \psi - \tilde{\eta}, \qquad \tilde{\eta}:= \eta + \A\psi -\psi\A.
\ee
The transformation rule for $\B$ then becomes
\be\label{delta-B}
\phi \delta \B = d_\A \tilde{\eta} - \F\psi +\psi \F.
\ee
After some transformations the rule for $c$ can be shown to become
\be\label{delta-c}
\delta c + \phi {\bf B} \delta {\bf A} = dh - {\rm Tr}(\tilde{\eta} {\bf F}).
\ee
It is easy to check that the Lagrangian (\ref{cdc}) is invariant under (\ref{delta-A}), (\ref{delta-B}) and (\ref{delta-c}). 

The transformation rule (\ref{delta-A}) in particular explains why (\ref{cdc}) is topological. Indeed, the $\tilde{\eta}$ part of this transformation tells us that any connection is gauge equivalent to any other connection in this theory. 

\subsection{Computation of the metric}

We now compute the metric (\ref{metric}) defined by $\Omega$. In the parametrisation (\ref{C}) we have
\be\label{met-1}
-6 g_\Omega(\xi,\eta) {\rm Vol}_\Omega / \left( - \frac{2}{3} {\rm Tr}(W^3)\right) = 
3\phi^5 ( -2{\rm Tr}(i_\xi W B)) ( -2{\rm Tr}(i_\eta W B)) 
\\ \nonumber
+3\phi^4 \left( ( -2{\rm Tr}(i_\xi W B)) i_\eta c+ ( -2{\rm Tr}(i_\eta W B)) i_\xi c\right) 
+\phi^3 4{\rm Tr}( i_\xi B i_\eta B B).
\ee
In deriving this result we have used the trace identities (\ref{trace-identities}). 
We have also used the fact that the contribution to the metric on the base
\be
i_\xi c\, i_\eta c=0.
\ee
This is easily checked by parametrising $c$ as the dual of a vector field. Then the wedge product in the above formula reduces to a single instance of the $\epsilon$-tensor with two insertions of the vector field, which is zero. 

To continue the calculation of the metric we need to parametrise
\be\label{B-param}
B^i =\sqrt{X}^{ij} \Sigma^j,
\ee
where $\Sigma^i$ are the anti-self-dual (ASD) 2-forms for the (conformal) metric defined by $B^i$ via Urbantke formula (\ref{urb-metric}). The matrix $X^{ij}$ is that of the wedge products 
\be
B^i B^j = - 2 X^{ij} {\rm vol}_\Sigma,
\ee
where ${\rm vol}_\Sigma$ is the volume form of the metric whose orthonormal ASD 2-forms are $\Sigma^i$. Explicitly $\Sigma^i \Sigma^i = - 6{\rm vol}_\Sigma$. For our sign conventions in the choice of $\Sigma$ see the next Section. We also need to parametrise $c$, and we do so via a vector field $v$ that is dual to the 3-form $c$
\be\label{c-param}
c= -2 ({\rm det}(X))^{1/4} i_v {\rm vol}_\Sigma,
\ee
where the prefactor is for future convenience. With these parametrisations we obtain the following metric
\be\label{met-2}
g_\Omega (\xi,\eta) {\rm Vol}_\Omega/ \left( - \frac{2}{3} {\rm Tr}(W^3) {\rm vol}_\Sigma\right) = \phi^5 i_\xi W^i X^{ij} i_\eta W^j \\ \nonumber
+ \phi^4 ({\rm det}(X))^{1/4} \left( i_\xi W^i \sqrt{X}^{ij} i_\eta (i_v \Sigma^j) +  i_\eta W^i \sqrt{X}^{ij} i_\xi (i_v \Sigma^j)\right)+
\phi^3 {\rm det}(\sqrt{X}) g(\xi,\eta)_\Sigma.
\ee
We now need to compute the determinant of the matrix appearing on the right-hand-side. This is done by noticing that the matrix of the quadratic form can be written as
\be
\left( \begin{array}{cc} \phi^{5/2} \sqrt{X} & 0 \\ 0 & \phi^{3/2} ({\rm det}(X))^{1/4} e_\Sigma \end{array}\right)
\left( \begin{array}{cc}\id & i_v \Sigma \\ i_v \Sigma & \id \end{array}\right)\left( \begin{array}{cc} \phi^{5/2} \sqrt{X} & 0 \\ 0 & \phi^{3/2} ({\rm det}(X))^{1/4} e_\Sigma \end{array}\right),
\ee
where $e_\Sigma\equiv e^I_\mu $ is the frame for the metric with ASD forms $\Sigma^i$, i.e. $g(\xi\eta)_\Sigma = i_\xi e^I i_\eta e^J \delta_{IJ}$, and $i_v \Sigma$ is the matrix $(i_v\Sigma)^i_I$. The factors of $({\rm det}(X))^{1/4}$ was introduced in (\ref{c-param}) precisely so that such a simple decomposition is possible. It is now easy to compute the determinant. We have
\be
{\rm det}\left( \begin{array}{cc}\id & i_v \Sigma \\ i_v \Sigma & \id \end{array}\right)= (1-|v|^2)^3. 
\ee
The other factor reads
\be
{\rm det}\left( \begin{array}{cc} \phi^{5} X & 0 \\ 0 & \phi^3 ({\rm det}(X))^{1/2} g_\Sigma \end{array}\right)
= \phi^{27} ({\rm det}(X))^3 {\rm det}(g_\Sigma).
\ee 
We now use
\be 
{\rm det}( g_\Omega {\rm vol}_\Omega) = ({\rm det}(g_\Omega))^{9/2}.
\ee
Thus, to compute ${\rm Vol}_\Omega=({\rm det}(g_\Omega))^{1/2}$ we need to take the power $1/9$ of the determinant of the quadratic form on the right-hand-side of (\ref{met-2}). This gives
\be\label{vol}
{\rm Vol}_\Omega = - \frac{2}{3} {\rm Tr}(W^3) {\rm vol}_\Sigma\, \phi^3 ({\rm det}(X))^{1/3} (1-|v|^2)^{1/3}.
\ee
This gives the metric
\be\label{met-3}
(1-|v|^2)^{1/3} g_\Omega(\xi,\eta) =  i_\xi W^i \frac{\phi^2 X^{ij}}{({\rm det}(X))^{1/3}} i_\eta W^j \\ \nonumber
\phi ({\rm det}(X))^{-1/12} \left( i_\xi W^i \sqrt{X}^{ij} i_\eta (i_v \Sigma^j) +  i_\eta W^i \sqrt{X}^{ij} i_\xi (i_v \Sigma^j)\right)+{\rm det}(\sqrt{X})^{1/6} g(\xi,\eta)_\Sigma.
\ee
We note that there is conformal freedom introduced with parametrisation (\ref{B-param}). This is the freedom of rescaling $g_\Sigma \to \Lambda^2 g_\Sigma$ and thus $\Sigma^i \to \Lambda^2 \Sigma^i$. This is to be done simultaneously with $\sqrt{X}^{ij}\to \Lambda^{-2} \sqrt{X}^{ij}$. We have parametrised $c$ in such a way that in order for this 3-form to be invariant under this parametrisation we need also to change $v\to \Lambda^{-1} v$. With this scaling the norm $|v|^2$ is invariant. It is also easy to see that the metric (\ref{met-3}) is invariant. 

The metric (\ref{met-3}) is not diagonal, which means that vector fields annihilated by $W^i$ are {\it not} the horizontal vector fields. The true horizontal vector fields are computed by searching for a vector field of the form
\be
\eta_H = \eta^\mu \left( \frac{\partial}{\partial x^\mu}\right) + \eta^\mu \alpha^i_\mu \left( \frac{\partial}{\partial W^i}\right),
\ee
where $\partial/\partial W^i$ are vector fields dual to one-forms $W^i$ (and orthogonal to the basic one-forms). Taking a product of such a vector field with a vertical vector field $\xi$ we get the following equation
\be
0= i_\xi W^i \frac{\phi^2 X^{ij}}{({\rm det}(X))^{1/3}} \alpha^j_\mu \eta^\mu + \phi ({\rm det}(X))^{-1/12} i_\xi W^i \sqrt{X}^{ij} v^\alpha \Sigma^j_{\alpha\mu} \eta^\mu.
\ee
This allows us to read off 
\be\label{conn-shift}
\alpha^i_\mu = - \phi^{-1} ({\rm det}(X))^{1/4} (X^{-1/2})^{ij} v^\alpha \Sigma^j_{\alpha\mu}.
\ee

The true metric on the base is now given as the contraction of two horizontal vector fields, where we need to take into account contributions (\ref{conn-shift}). The result is
\be\label{metric-M}
(1-|v|^2)^{1/3} g_\Omega(\xi_H,\eta_H) = ({\rm det}(X))^{1/6} \left( (1-|v|^2) g_\Sigma(\xi,\eta) + g_\Sigma(v,\xi) g_\Sigma(v,\eta)\right).
\ee
Note that this is {\it not} a conformal rescaling of the metric $g_\Sigma$. In computing this we have used the identity
\be
\Sigma^i_{\alpha\mu} \Sigma^i_{\beta\nu} = g_{\alpha\beta} g_{\mu\nu} - g_{\alpha\nu} g_{\mu\beta} - \epsilon_{\alpha\mu\beta\nu}.
\ee
For completeness, the contraction of two vertical vector fields is
\be
(1-|v|^2)^{1/3} g_\Omega(\xi_V,\eta_V) = \xi^i \frac{\phi^2 X^{ij}}{({\rm det}(X))^{1/3}}\eta^j.
\ee
As a check, now that the metric is written in a block-diagonal form, we can compute the determinant. The determinant of the vertical-vertical block is $\phi^6/(1-|v|^2)$. The determinant of the horizontal-horizontal block can be computed by pointing $v$ along one the axes, e.g. the first. The metric then becomes 
\be
({\rm det}(X))^{1/6} {\rm diag}\left( (1-|v|^2)^{-1/3}, (1-|v|^2)^{2/3},(1-|v|^2)^{2/3},(1-|v|^2)^{2/3}\right).
\ee
The determinant of the horizontal-horizontal part is then $({\rm det}(X))^{2/3} (1-|v|^2)^{5/3}$. The product of two block determinants is $\phi^6 ({\rm det}(X))^{2/3} (1-|v|^2)^{2/3}$. The square root of the determinant is then $\phi^3 ({\rm det}(X))^{1/3} (1-|v|^2)^{1/3}$, which agrees with what we previously computed. 

\subsection{Dimensionally reduced action}

Putting together (\ref{cdc}) and (\ref{vol}) we get the dimensionally reduced action
\be\label{action-dim-red}
S[\B,\A,\phi,v]=\int_M -2\, {\rm Tr}(\phi^4 {\bf B}{\bf F} + (\phi^2/2) {\bf B}{\bf B}) + 6 \phi^2 ({\rm det}(X))^{1/4} (v^\mu \partial_\mu \phi ) {\rm vol}_\Sigma
\\ \nonumber
+ 3\lambda  \phi^3 ({\rm det}(X))^{1/3} (1-|v|^2)^{1/3} {\rm vol}_\Sigma \, .
\ee

\subsection{Interpretation}

To give interpretation to this action, we first note that if one sets $\phi=const$ one obtains a specific theory from the class of "Deformations of General Relativity" studied by the present author in a series of works starting with \cite{Krasnov:2006du}. A particularly relevant reference is \cite{Krasnov:2008fm}, where it is shown that ${\rm SU}(2)$ BF theory with a general potential for the $B$-field is a gravity theory with two propagating degrees of freedom. The reference \cite{Krasnov:2009ik} explains how these BF-type gravity theories can be explicitly recast into metric form. The main idea is to parametrise the $B$-field by a metric it determines as well as a set of auxiliary scalar fields. The scalar fields are non-dynamical and can be eliminated from the action by solving their field equations. This gives gravitational Lagrangians starting with the Einstein-Hilbert term, but corrected with an infinite number of higher powers of the curvature terms, see \cite{Krasnov:2009ik} for details of this procedure. 

A very interesting feature of (\ref{action-dim-red}) is that for $\phi=const$ the value of the effective 4D cosmological constant is determined by $\phi$, see \cite{Krasnov:2016wvc} for more details on this aspect of the dimensionally reduced theory. Moreover, as is shown in \cite{Krasnov:2016wvc}, for values $\lambda\phi\approx 1$ the 4D cosmological constant is arbitrarily small {\it and} the deviations of the gravity theory (\ref{action-dim-red}) from General Relativity for curvatures smaller than Planckian are negligible. 

The new feature of the action (\ref{action-dim-red}) is that there is also a scalar field on top of the ${\bf B}, {\bf A}$ fields present in the theories studied in \cite{Krasnov:2008fm}, \cite{Krasnov:2009ik}. Following the same steps as in \cite{Krasnov:2009ik} one can envisage eliminating from the Lagrangian all fields apart from the metric and the scalar field $\phi$, and obtaining a scalar-tensor theory of a specific type. 

Prior to eliminating any fields, the action (\ref{action-dim-red}) is first-order in derivatives. In particular, it is clear that the vector field $v^\mu$ is an auxiliary field needed to put the second-order scalar field Lagrangian into a first-order form. Thus, note that the Euler-Lagrange equation for $v^\mu$ that follows from (\ref{action-dim-red}) is an algebraic equation for $v^\mu$ in terms of the derivative $\partial_\mu \phi$. Solving this equation, while difficult explicitly because of the presence of the cubic root, is possible in principle. Eliminating $v^\mu$ in this fashion, one obtains the Lagrangian for $\phi$ of the type 
\be
{\cal L} = K( \phi^3, |\partial_\mu \phi^3|^2).
\ee
This type of scalar theories has been studied under the name of "K-essence" \cite{ArmendarizPicon:1999rj}.

Another important point about the scalar-tensor theory under discussion is that the metric (\ref{metric-M}) that this theory describes, i.e. the metric that gets induced by $\Omega$ on the 4D base $M$, is of the form
\be
g^{phys}_{\mu\nu} = f_1( \phi, v) g_{\mu\nu} + f_2(\phi, v) v_\mu v_\nu,
\ee
where we denoted by $g^{phys}_{\mu\nu}$ the metric induced by $\Omega$, and $g_{\mu\nu}$ is the metric determined by ${\bf B}$. Theories with this type of dependence of the physical metric on the scalar field have also been studied in the literature on scalar-tensor theories. 

In the next section we use the formalism developed above to show how some very symmetric solutions of the equations (\ref{feqs}) can be obtained.

\section{Some solutions}
\label{sec:spheres}

The goal of this section is to explicitly determine some solutions to (\ref{feqs}), assuming ${\rm SU}(2)$ invariance as in the previous section. Thus, we will be working with the 4D theory (\ref{action-dim-red}) and its field equations. The simplest solutions of a 4D gravity theory are cosmological ones, and so we will make the assumption that fields on the 4D base are homogeneous isotropic. Our aim is to exhibit the squashed and round $S^7$ solutions of the theory (\ref{action}). Both solutions to be described can be viewed as $S^3$ bundles over $S^4$, and this is why the reduction on $S^3$ formalism developed in the previous section is relevant. 

\subsection{Homogeneous isotropic 3-forms}

A 3-form on $X$ that is ${\rm SU}(2)$ invariant along the 3D fibers and is homogeneous isotropic along the 4D base can be parametrised as follows
\be\label{omega-str}
\Omega = -2\, {\rm Tr} \left( \frac{1}{3}\phi^3 W^3 + \phi W \Sigma\right) + c,
\ee
Here $c$ is a 3-form on the base, which can only be (due to symmetry) a multiple of $\e^3$
\be
c= -2\xi \gamma^3 \left( -\frac{2}{3} {\rm Tr}(\e^3)\right).
\ee
This is the same parametrisation as (\ref{c-param}), with $v= (\xi/\beta) (\partial/\partial t)$ so that the norm of $v$ computed using the metric 
\be\label{metric-4d}
ds^2 = \beta^2 dt^2 + \gamma^2 \sum_i (e^i)^2
\ee 
is $|v|^2=\xi^2$. All other objects are as follows
\be\label{Sigma}
W:= \m+ \alpha \e, \qquad \Sigma :=  \beta\gamma dt \e - \gamma^2 \e^2,
\ee
where $\alpha,\beta,\gamma$ are function of time. Note that the Urbantke metric determined by $\Sigma$ is (\ref{metric-4d}). Thus, we think of the 4D base as $\R\times S^3$ are $\e$ are the usual Lie algebra valued 1-forms on $S^3$ satisfying
\be
d\e = 2\kappa \e^2.
\ee
The curvature of ${\bf A}=\alpha\e$ is then
\be
\F= \alpha' dt \e +\alpha(\alpha+2\kappa) \e^2.
\ee

\subsection{The action}

Using $-(2/3){\rm Tr}(\e^3) = e^3\equiv (1/6)\epsilon^{ijk} e^ie^j e^k$ we have
\be
-2{\rm Tr}(\Sigma {\bf F}) = 3dt e^3 ( - \alpha' \gamma^2 + \alpha(\alpha+2\kappa) \beta \gamma), \\ \nonumber
-2{\rm Tr}(\Sigma \Sigma) = - 6 dt e^3 \beta\gamma^3.
\ee
The dimensionally reduced action (\ref{action-dim-red}) is then
\be\label{L-red}
S/3= \int dt \Big[ -\phi^4\gamma^2 \alpha' + 2\phi^2 \gamma^3 \xi \phi' 
\\ \nonumber
+ \alpha(\alpha +2\kappa) \phi^4 \beta\gamma - \phi^2 \beta\gamma^3 + \lambda \phi^3 \beta \gamma^3 (1 -\xi^2)^{1/3} \Big],
\ee
where we omitted the unimportant spatial volume factor. 

Varying this action with respect to the fields, we get a set of equations. It is convenient to solve these equations for the first derivatives of the fields. We get
\be\label{evol-eqs}
\alpha'= \frac{\lambda \beta \gamma\xi^2}{(1-\xi^2)^{1/3} \phi^2} - \frac{\beta\alpha(\alpha+2\kappa)}{\gamma}, \qquad
\phi' = \frac{\lambda \beta \xi\phi}{3(1-\xi^2)^{2/3}}, \\ \nonumber
\gamma' = -\beta (\alpha+\kappa) - \frac{2\lambda \beta \gamma \xi}{3(1-\xi^2)^{2/3}}, \qquad
\xi' = \frac{\beta}{2\phi}\left( 1 + 6(\alpha+\kappa) \xi \frac{\phi}{\gamma} + 5 \alpha(\alpha+2\kappa) \frac{\phi^2}{\gamma^2}\right).
\ee
Here to simplify the first and the last equations we have used the constraint
\be\label{constr}
\lambda\phi(1-\xi^2)^{1/3} = 1 - \alpha(\alpha+2\kappa) \frac{\phi^2}{\gamma^2},
\ee
which is obtained by varying the action with respect to $\beta$ and dividing by $\gamma^3 \phi^2$.

\subsection{The metric}

For the 3-form (\ref{omega-str}) the metric (\ref{met-3}) becomes
\be
(1-\xi^2)^{1/3} ds^2_\Omega = \phi^2 (W^i)^2 + \phi \gamma\xi (W^i e^i + e^i W^i) + \beta^2 dt^2 + \gamma^2 \sum_i (e^i)^2.
\ee
This is not block diagonal, but diagonalises if we write it as
\be
ds^2_\Omega =  \frac{\phi^2}{(1-\xi^2)^{1/3}} \left(W^i+ \frac{\gamma\xi}{\phi} e^i \right)^2  + (1-\xi^2)^{-1/3} \beta^2 dt^2 + (1-\xi^2)^{2/3} \gamma^2 \sum_i (e^i)^2.
\ee

\subsection{Field redefinition}

By looking at the metric (\ref{metric-spheres}) we see that it makes sense to set 
\be
\beta^2 = (1-\xi^2)^{1/3}
\ee
and introduce quantities
\be
\tilde{\phi}^2:= \frac{\phi^2}{(1-\xi^2)^{1/3}}, \qquad \tilde{\alpha} = \alpha + \frac{\gamma\xi}{\phi}, \qquad \tilde{\gamma} = (1-\xi^2)^{1/3}\gamma.
\ee
It is also convenient to introduce a new variable $\theta$ so that
\be
\xi= \sin(\theta).
\ee

\subsection{Lagrangian and equations after field redefinition}

The most efficient way of obtaining the field equations for the tilded quantities is to carry out the field redefinition in the Lagrangian (\ref{L-red}), and then derive the new evolution equations. We will now omit tildes from all the quantities, and give the Lagrangian in terms of the new variables. We get
\be\label{action-reduced}
S/6 = \int dt \Big[ \frac{2}{3} \gamma^3\phi^3 \theta' - \phi^4 \gamma^2 \alpha'\\ \nonumber
+\phi^2\beta\gamma\left(  \cos(\theta)(-\gamma^2+\phi^2\alpha(\alpha+2\kappa)) - 2\sin(\theta) \phi \gamma (\alpha+\kappa) \right)  +\lambda \phi^3 \beta\gamma^3 \Big].
\ee
We get the following Euler-Lagrange equations from this Lagrangian, rewritten as a set of equations for the first derivatives of the fields
\be
\gamma' = -(\alpha+\kappa) \cos(\theta) -\beta \sin(\theta) \alpha (\alpha+2\kappa) \frac{\phi}{\gamma}, \qquad
\phi' = \frac{1}{2} \sin(\theta)\left(1+\alpha (\alpha+2\kappa)\frac{\phi^2}{\gamma^2}\right), \\ \nonumber
\alpha' = \frac{\gamma}{2\phi^2 } \left( \cos(\theta)(1+ 3\alpha(\alpha+2\kappa)\frac{\phi^2}{\gamma^2}) - 2\sin(\theta) (\alpha+\kappa) \frac{\phi}{\gamma} \right), \\ \nonumber
\theta' = \frac{1}{2\phi} \left( \cos(\theta)(1+ 5\alpha(\alpha+2\kappa)\frac{\phi^2}{\gamma^2}) - 4\sin(\theta) (\alpha+\kappa) \frac{\phi}{\gamma} \right).
\ee
The constraint becomes
\be\label{constr*}
\lambda\phi = \cos(\theta)(1- \alpha(\alpha+2\kappa)\frac{\phi^2}{\gamma^2}) +2\sin(\theta) (\alpha+\kappa) \frac{\phi}{\gamma} .
\ee

In terms of the new variables the metric is
\be\label{metric-spheres}
ds^2_\Omega = \phi^2 (m^i + \alpha e^i)^2 + dt^2 + \gamma^2 \sum_i (e^i)^2.
\ee

\subsection{Squashed $S^7$}

The simplest solution of the above system is the so-called squashed $S^7$. This is obtained by demanding $\phi=const, \theta=0$. The condition that $\theta'=0$ then implies that we must have
\be\label{squashed-1}
\alpha(\alpha+2\kappa) \frac{\phi^2}{\gamma^2}=-\frac{1}{5}.
\ee
We also have the equations for $\gamma$ and $\alpha$
\be
\gamma' = - (\alpha+\kappa), \qquad \alpha' = \frac{\gamma}{5\phi^2}.
\ee
A solution of this system is
\be
\gamma = \cos\left(\frac{t}{\phi\sqrt{5}}\right), \qquad \alpha+\kappa =  \frac{1}{\phi\sqrt{5}} \sin\left(\frac{t}{\phi\sqrt{5}}\right)
\ee
Then (\ref{squashed-1}) is solved by
\be
\kappa^2 = \frac{1}{5\phi^2}.
\ee
We also have the constraint that gives
\be
\lambda\phi = \frac{6}{5}.
\ee
If we fix the normalisation of the size of the spatial $S^3$ by choosing $\kappa=1$ we get
\be
\phi = \frac{1}{\sqrt{5}}
\ee
and the metric (\ref{metric-spheres}) becomes
\be\label{squashed}
ds^2_{squashed} = \frac{1}{5} (m^i - (1- \sin(t)) e^i)^2 + dt^2 + \cos^2(t) \sum_i (e^i)^2.
\ee
The presence of $1/5$ in front of the fiber metric explains the terminology "squashed" $S^7$. This is the metric on the total space of an $S^3$ fibration over $S^4$.

\subsection{Round $S^7$}

Another solution with $\phi=const$ is obtained by setting to zero the term in the brackets in the equation for $\phi'$
\be\label{round-2}
\alpha (\alpha+2\kappa)\frac{\phi^2}{\gamma^2}=-1.
\ee
In this case the equations for $\alpha',\theta'$ can be simplified to
\be\label{round-1}
\alpha' = -\frac{\gamma}{\phi^2 } \left( \cos(\theta)+ \sin(\theta) (\alpha+\kappa) \frac{\phi}{\gamma}\right), \\ \nonumber
\theta' = -\frac{2}{\phi} \left( \cos(\theta)+ \sin(\theta) (\alpha+\kappa) \frac{\phi}{\gamma}\right).
\ee
But also the constraint simplifies and becomes
\be\label{round-4}
\frac{\lambda\phi }{2} = \cos(\theta)+ \sin(\theta) (\alpha+\kappa) \frac{\phi}{\gamma}.
\ee
So, we can use this constraint to simplify the equations (\ref{round-1})
\be\label{round-3}
\alpha' = -\frac{\lambda\gamma}{2\phi }, \qquad \theta' = -\lambda.
\ee
The second equation here gives 
\be
\theta = -\lambda t,
\ee
where a choice of the origin of time was made to eliminate the integration constant. The equation for $\alpha'$ can be solved by expressing $\alpha+\kappa$ from (\ref{round-2}) as
\be\label{round-5}
\frac{\gamma}{\phi}= \sqrt{\kappa^2-(\alpha+\kappa)^2}.
\ee 
Substituting this into the first equation in (\ref{round-3}) we get a closed equation for $\alpha+\kappa$, with the solution
\be
\alpha+\kappa = - \kappa \sin(\lambda t/2).
\ee
The choice of integration constant here was made so that when this is substituted into the constraint (\ref{round-4}) we have $\phi=const$. Indeed, this substitution gives
\be\label{round-6}
\lambda\phi=2.
\ee
As the last check, we note that the equation for $\gamma'$ becomes
\be
\gamma' =\kappa\left(  \sin(\lambda t/2) \cos(\lambda t) -  \cos(\lambda t/2) \sin(\lambda t) \right) = - \kappa \sin(\lambda t/2) .
\ee
This is compatible with the solution (\ref{round-5}) provided (\ref{round-6}) holds. 

If we choose $\lambda=2, \kappa=1$ we get $\phi=1$ and the following metric
\be
ds^2_{round} = (m^i - (1+ \sin(t)) e^i)^2 + dt^2 + \cos^2(t) \sum_i (e^i)^2,
\ee
which is essentially the same as (\ref{squashed}) but without the $1/5$ in front of the first term. It can be shown by an explicit computation (using quaternionic Hopf projection) that this is the standard round metric on the $S^7$, here described as an $S^3$ fibration over $S^4$. 

\section{Discussion}

In this paper we studied a simple dynamical theory (\ref{action}) of 3-forms in 7 dimensions. We have characterised this theory in two ways. First, we used the $6+1$ decomposition and demonstrated that the phase space of the theory is the space of 3-forms on the "spatial" 6D slice, and that there are 7 first class constraints. These constraints are just those generating the spatial and temporal diffeomorphisms. The dimension of the reduced configuration space is 3. Second, we performed the dimensional reduction of the theory (\ref{action}) to 4D, assuming ${\rm SU}(2)$ invariance of the 3-form. We obtained (\ref{action-dim-red}), which is a variant of scalar-tensor theory in 4 dimensions. We have then used the ${\rm SU}(2)$-invariant ansatz, together with a further assumption about homogeneity-isotropy on the 4D slice, to exhibit some simple solutions of the field equations (\ref{feqs}). 

An incomplete list of open questions is as follows. First, it would be very interesting to study the theory (\ref{action}) quantum mechanically, by attempting to compute (perturbatively) the path integral. It is easy to see that the theory is power-counting non-renormalisable. The form of the kinetic term $\Omega d\Omega$ tells us that we should give the field $\Omega$ the mass dimension 3. The interactions then start with $\Omega^3$, and thus have a coupling constant of mass dimension $-2$. 

To study the quantum theory perturbatively, we would need to expand around a non-trivial $\Omega$ background, because the presence of a root in the potential ${\rm Vol}_\Omega$ term makes the expansion of this term only well-defined around a non-zero $\Omega$. At one loop, simple power counting shows that the self-energy diagram can diverge as the 5th power of the momentum. Thus, at one loop, one should expect the leading divergence to be of the schematic type 
\be
\frac{1}{M^4} \Delta^2 \Omega d\Omega,
\ee
where $1/M^2$ is the coupling constant in front of the cubic interaction, and $\Delta$ is some appropriate Laplace-type operator. Experience with the one loop behaviour of GR suggests that this divergence, if at all present, may be removable by a field redefinition of the type 
\be
\Omega \to \Omega + \frac{1}{M^4} \Delta^2 \Omega.
\ee
It would be very interesting to compute the one-loop effective action for (\ref{action}) and see if any divergences remain after renormalisation by field redefinitions. The natural conjecture is that at most the constant in front of the action gets one-loop renormalised. 

A harder calculation is that of divergences at two loops. This calculation would be particularly interesting given the fact that we are dealing with a theory of 3-forms, and so it is not easy to come up with a candidate two-loop counterterm if it is to be written in terms of differential forms. It would be very interesting to perform the 2-loop computation, as it would significantly improve our intuition on the quantum behaviour of power-counting non-renormalisable diffeomorphism invariant theories with propagating degrees of freedom. The only available example of such a calculation is that \cite{Goroff:1985th} for 4D GR. The difference with the calculation envisaged here is that 4D GR is a theory of metrics, which makes it easier to write potential counterterms. 

The other set of open questions relates to a possible physical interpretation of the theory (\ref{action}). As it stands, this theory should be interpreted as a purely gravitational theory, with possibly an extra scalar. There are clearly no matter degrees of freedom described by (\ref{action}), definitely no fermionic degrees of freedom. Thus, if this set of ideas is ever to be developed into a physical theory, one must define how other known bosonic fields (e.g. gauge fields) and fermions couple to this type of gravity. The fact that it seems to be possible to describe gravity with differential forms suggests that one should try to use the same formalism for describing all other building blocks of Nature. It remains to be seen how far this idea can be pushed. 

The other open question is whether the theory (\ref{action}) reduces to General Relativity in some regime. As we have indicated in the main text, and as reference \cite{Krasnov:2016wvc} discusses in much more details, it is possible to get a 4D theory that is arbitrarily close to General Relativity by dimensionally reducing (\ref{action}) on $S^3$ of a fixed size $\phi=const$, and tuning this constant appropriately. However, as is the case also with more familiar Kaluza-Klein theories, see e.g. \cite{Duff:1986hr}, it is probably inconsistent to just freeze this scalar degree of freedom by hand. Instead, the right approach should be to allow this field to be dynamical, and let it settle dynamically to some value. However, as we have seen in Section \ref{sec:spheres}, the natural values are $\lambda\phi=5/6$ for the squashed sphere and $\lambda\phi=2$ for the round sphere solutions. Neither of this is the value $\lambda\phi=1$ that would give an approximately flat 4D base. And it is also intuitively clear that if one allows $\phi$ to be dynamical so that in particular the 7D metric defined by $\Omega$ is Einstein, it is impossible to have the internal $S^3$ strongly curved (and thus small) while the base is weakly curved (and large). So, there appears to be no solution of the full set of equations of the theory (\ref{action}) that approximates General Relativity. But given that the difficulty of explaining what tunes the size of the extra dimensions to a phenomenologically acceptable value is also shared by the more traditional Kaluza-Klein theories, we feel that further study is necessary before a definite conclusion on this set of ideas is reached.  

The other pressing question about the formalism developed in this paper is how to describe a 4D world with a Lorentzian signature metric. The solutions that we have described in Section \ref{sec:spheres} suggest that the most appropriate way to do this is to analytically continue the $t$ coordinate in (\ref{Sigma}) to imaginary values. This will make the 3-form $\Omega$ complex, but keep the metric on the 4D base, as well as in the 3D fibers, real. For the particular solutions we described the 4D metric becomes that of de Sitter space. The price one pays for this is that the off-diagonal components of the 7D metric become complex. This may be physically acceptable given that the off-diagonal components of the 7D metric just encode a certain 4D connection field, and one may allow this to be complex. In fact, the connection field in question has the interpretation of the self-dual part of the Levi-Civita connection, and this is complex in Lorentzian signature. So, this complexity of the 7D metric may not be a problem. We will not attempt to develop a "Lorentzian" version of the theory in the present paper, leaving this subtle problem to a different publication. 

\section*{Acknowledgments}

The author was supported by ERC Starting Grant 277570-DIGT. The author is grateful to the Max-Planck-Institute for Gravitational Physics (Albert Einstein Institute), Golm (Potsdam) for hospitality while this work has been carried out. I am grateful to Yuri Shtanov and Yannick Herfray for numerous discussions on the subject of this paper. 

\section*{Appendix}

\subsection*{Notations}

The common notation is that a bold face letter denotes an object that is Lie algebra valued. These can be expanded in generators. Thus, e.g. ${\bf e} = \tau^i e^i$, where the
objects $\tau^i :=-(\im/2)\sigma^i$ are usual generators of ${\mathfrak su}(2)$. We have
\be
\tau^i \tau^j = -\frac{1}{4} \delta^{ij} \id + \frac{1}{2} \epsilon^{ijk} \tau^k
\ee
so that in particular $\epsilon^{ijk}\tau^j\tau^k = \tau^i$. The trace everywhere is the usual matrix trace. 

\subsection*{Canonical expression for the 3-form}

The canonical expression for the 3-form that gives a Riemannian metric is
\be\label{omega-can}
\Omega = e^{567} +  e^5 (e^{41}-e^{23}) +  e^6( e^{42}- e^{31}) +  e^7 (e^{43}- e^{12}),
\ee
where we used the usual notation $e^{ij\ldots k}=e^i e^j \ldots e^k$, and again the wedge product is implied. The metric obtained via (\ref{metric}) is the metric of Riemannian signature 
\be
ds^2 = \sum_{a=1}^7 (e^a)^2.
\ee 
The form (\ref{omega-can}) can also be written more compactly as
\be
\Omega =  e^{567} + e^5 \Sigma^1 + e^6 \Sigma^2 + e^7 \Sigma^3,
\ee
where
\be
\Sigma^1 = e^{41}-e^{23}, \quad \Sigma^2 =   e^{42}- e^{31}, \quad \Sigma^3 =  e^{43}- e^{12}
\ee
are the basic ASD 2-forms. Let us also give the expression for the dual 4-form
\be\label{omega-dual}
{}^*\Omega =  e^{4123} - \Sigma^1 e^{67} - \Sigma^2 e^{75} - \Sigma^3 e^{56}.
\ee

\subsection*{Relations between field equations}

Not all equations in (\ref{feqs}) are independent, and our goal here is to state the corresponding identities. Indeed, because of diffeomorphism invariance, there must be some relations between the field equations. To derive these relations we use the invariance of the action (\ref{action}) with respect to diffeomorphisms. We have, for the variation of the action
\be
\delta S[\Omega]=2\int \delta\Omega \left( d\Omega - \lambda {}^*\Omega\right).
\ee
We now substitute here
\be
\delta \Omega = {\cal L}_\xi \Omega = d(i_\xi \Omega) + i_\xi d\Omega.
\ee
The action is diffeomorphism invariant so we must have
\be\label{S-diff}
0= \int d(i_\xi \Omega)(d\Omega -\lambda {}^*\Omega) + i_\xi d\Omega (d\Omega - \lambda {}^*\Omega).
\ee
We have a simple identity
\be
0=i_\xi (d\Omega d\Omega)=i_\xi d\Omega \,d\Omega + d\Omega \,i_\xi d\Omega = 2i_\xi d\Omega \,d\Omega,
\ee
and so the third term in (\ref{S-diff}) is zero. Similarly
\be\label{s-diff-1}
0=i_\xi (d\Omega {}^*\Omega) = i_\xi d\Omega \,{}^*\Omega + d\Omega\, i_\xi {}^*\Omega.
\ee
Integrating by parts in the first two terms in (\ref{S-diff}), and applying (\ref{s-diff-1}) to the fourth term we get
\be
0=\lambda \int \left( i_\xi \Omega \,d{}^*\Omega + d\Omega \,i_\xi {}^*\Omega\right),
\ee
which is only possible if
\be\label{identity}
i_\xi \Omega \,d{}^*\Omega + d\Omega \,i_\xi {}^*\Omega=0.
\ee
This identity must hold automatically, as a consequence of the definition of ${}^*\Omega$. This identity can also be interpreted as giving a set of relations between the field equations (\ref{feqs}) and their first derivatives (\ref{d-feqs}). Indeed, multiplying the left hand-side $d\Omega -\lambda {}^*\Omega$ of the field equations by $i_\xi {}^*\Omega$ we get the second term in (\ref{identity}). Multiplying the left-hand-side $d{}^*\Omega$ of (\ref{d-feqs}) by $i_\xi\Omega$ we get the first term in (\ref{identity}). Thus, there exists a linear combination of the field equations and their first derivatives that is identically zero. The field equations are therefore not all independent, as must be the case in a theory that is diffeomorphism invariant.

\end{document}